\documentclass[aps,prd,twocolumn,groupedaddress,showpacs,nofootinbib,amssymb,balancelastpage]{revtex4}
\usepackage[dvipdfmx]{graphicx}
\usepackage{graphicx,bm,color}
\usepackage{amsmath}
\usepackage{amssymb}
\usepackage{amsfonts}
\usepackage{cases}
\usepackage{cancel}

\usepackage{xcolor}
\definecolor{orange}{RGB}{255,127,0}

\begin{document}

\title{{\boldmath 
Enhanced diHiggs signal from hidden scalar QCD \\
at leading-order scale-symmetry limit}
\vspace{5mm}}


\author{Ruiwen Ouyang\footnote{ruiwen.ouyang@kbfi.ee}}
\affiliation{Center for Theoretical Physics and College of Physics, Jilin University, Changchun, 130012, China.}
\affiliation{Laboratory of High Energy and Computational Physics, National Institute of Chemical Physics and Biophysics, Ravala pst. 10, 10143 Tallinn, Estonia}
\author{Shinya Matsuzaki\footnote{synya@jlu.edu.cn}}
\affiliation{Center for Theoretical Physics and College of Physics, Jilin University, Changchun, 130012, China.}
\affiliation{Department of Physics, Nagoya University, Nagoya 464-8602, Japan}
\date{\today}


\begin{abstract}
We develop an effective-model description arising from a recently proposed scale-invariant hidden scalar-QCD, which has been used to explain the dynamical origin of the electroweak scale. In addition to the previous works, our new effective model includes the dynamical scale-anomaly effect from the hidden QCD gluons, to explicitly break the classical-scale invariance at the level of an effective field theory, which is known as the leading-order scale-symmetry (LOSS). In the phenomenological analysis, the proposed model predicts a light composite dilaton composed of hidden scalar quarks and gluons with the mass around electroweak scale (around 280 GeV), and has only one input parameter, which is the mixing angle between the Higgs boson and the composite dilaton. Our result for the dilaton mass is in accord with the lattice simulation for scalar QCD, where the scalar-quark bound states acquire a large effective mass from the hidden gluon contribution. Furthermore, we predict several significant deviations from the SM, like the diHiggs production cross sections (maximally about 10 times larger than the SM prediction), that could be directly tested at the high luminosity LHC. It is also the first study for the diHiggs production signal predicted from a scale(conformal)-invariant hidden sector, even from dark/hidden QCD. 
Our proposed effective model is thus significantly different than the conventional realization of scale-invariant hidden-scalar QCD without the scale anomaly effect, and can potentially provide a competitive explanation for many exotic phenomena beyond the standard model, such as new dark matter candidates and a strongly first-order electroweak phase transition.
\end{abstract}

\maketitle

\section{Introduction}
It is no doubt that the standard model (SM) meets perfectly with nearly all of experimental results in high-energy physics. However, from a theoretical point of view, the critical part in construction of the SM seems to be artificial: 
a negative mass-squared term for the Higgs field is introduced by hand to generate a nonzero vacuum expectation value that finally breaks the electroweak (EW) symmetry. Thus, it is still natural to ask ``what is the most fundamental nature behind this artifact?''  Moreover, if the SM is supposed to be a fundamental theory up to the Planck scale, then an extremely large quadratic divergence on the Higgs mass will arise from the quantum correction, that eventually produce a highly-accurate fine tuning for the Higgs mass, which is known as the fine-tuning problem regarding the EW symmetry breaking. Therefore, it would be suspected  
that a more fundamental theory should exist beyond the SM, which should explain the genesis of the EW scale in a more reasonable way to avoid such a fine tuning problem. 

On the way to going beyond the SM, conventionally we may start from what we have confirmed in the experiment, say, by taking into account the existence of the dark matter (DM), the observation of the neutrino oscillation, and the presence of several CP violations, etc. Meanwhile, we may also get some indirect hints from the currently ongoing experiments: no direct evidence has been found. That would imply null direct correlation between the SM and a new physics in terms of the current observational sensitivity, hence would urge us to consider a scenario with scale-invariance.

A recently proposed hidden scalar QCD \cite{Kubo:2015cna} is one of such kind of scenarios, which is build with the classical-scale invariance incorporated as a solution for the fine-tuning problem~\footnote{Strictly speaking, 
the classical-scale invariance may not be a solution to the gauge hierarchy problem. 
This kind of scenario would need to invoke a very stringent assumption about the UV physics: no hard thresholds should arise even including gravity. For detailed discussion about such assumptions and the quantum gravity, for example, see references~\cite{Shaposhnikov:2009pv,Wetterich:2016uxm}, on that the original scale-invariant scalar QCD model~\cite{Kubo:2015cna} we presently quote is based. 
Though it is still controversial,  
it would be worth exploring the phenomenology if the classical- scale invariance could somehow be realized (by coincidence) at the low-energy scale. That is our point of view on a theoretical ground for what we try to investigate 
through the present study.}
and allows only the Higgs to couple to the hidden QCD.    
In the typical scenario of hidden scalar QCD, 
the Higgs mass term is replaced by a Higgs portal coupling to a singlet scalar field $(S)$ 
having the hidden QCD charge, 
so that the EW-symmetry breaking scale can be dynamically generated from 
a nonperturbative condensation of $S^\dag S$ through the Higgs portal term~\cite{Kubo:2015cna}.    
Therefore, it is possible to explain the scalegenesis of the EW symmetry by 
the hidden scalar QCD. 

What's more, the hidden scalar QCD also naturally includes DM candidates 
because of the possible existence of stable scalar hadron states~\cite{Kubo:2015cna,Kubo:2017wbv}. 
Furthermore, it is also possible to realize a strongly-first order EW phase transition and a detectable primordial gravitational wave signals, as long as the Higgs portal coupling can be large enough~\cite{Kubo:2015joa,Kubo:2016kpb}. 

Thus, the hidden scalar QCD possesses a potential to explain 
the dynamical origin of the EW scalegenesis 
and provides rich related phenomenological consequences 
accessible in the future experiments. 

However, this scenario may still lack important ingredients.  
To capture such a possibly missing point, we may make use of the powerful tools about the effective theory of fermionic QCD. 
In the fermionic QCD case, it is believed that the gluon condensation effect arising from 
the nonperturbative-scale anomaly can be important for the low-energy effective theory of QCD, especially for the origin of hadron mass. 
Recently, based on this consideration, a chiral-scale effective theory has been built~\cite{Li:2016uzn}, where the scale anomaly was introduced as an essential effect to generate hadron masses. Combined with the classical-scale invariance, this minimal inclusion of the scale anomaly is known as the leading-order scale-symmetry (LOSS).

In this paper, we propose an effective model for hidden scalar QCD at the LOSS limit with the gluonic effect incorporated as the nonperturbative-scale anomaly. Our estimation about the gluonic contribution in the mass of hidden-scalar QCD meson may be consistent with the lattice simulations~\cite{Iida:2006fe,Iida:2007qp,Iida:2008cq}, as well as some nonperturbative analysis~\cite{Imai:2014yxa,Imai:2014tea}, about the estimation of dynamical mass of diquark bound state in the ordinary QCD, in which the diquark can play a role of a colored
scalar. 
It is demonstrated that the proposed model based on the LOSS gives a definite prediction to diHiggs signals at the LHC, in correlation with a significant deviation from the SM on the Higgs coupling measurement. 
This result is manifestly a smoking-gun of the LOSS limit, which will open a new avenue for the phenomenological probe of the hidden scalar QCD scenario, other than those so far explored, such as DM physics, a strongly-first order EW phase transition, and gravitational waves detectability.

This paper is structured as follows. In Sec.~II, we stat from a brief review on a classical-scale invariant-hidden scalar QCD sector and its possible low-energy description governed by low-lying scalar QCD hadrons in part A. Thereafter in part B we claim the importance of the inclusion of nonperturbative (gluonic-) scale anomaly, and the importance to introduce the LOSS limit. Then in part C the light composite dilaton with the mass around 280 GeV is shown to be a natural consequence of the hidden-scalar QCD at the LOSS limit. In Sec.~III we make several phenomenological predictions from our proposed model, such as the decay properties of the predicted light composite dilaton (in part A), the effects on the Higgs trilinear coupling (in part B), the estimation on the diHiggs production cross section (in part C), as well as the estimation on other dilaton-resonant channels like the sensitive resonant di-EW boson production cross section (in part D). Finally, Sec.~IV is devoted to summary of this paper and prospect on the LOSS for some future researches on the DM, as well as an issue on the EW phase transition.    

\section{Scale-Invariant Hidden-Scalar QCD in the LOSS limit}
Let us begin with a review of a classically-scale invariant-hidden scalar QCD. As in the literature~\cite{Kubo:2015cna}, we assume that there exist some SM-singlet scalars ($S_i ^a$), which are strongly coupled via a hidden scalar QCD gauge interaction, with the hidden flavors $i=1,2,...,N_{f}$ as well as the hidden QCD charges $a=1,2,...,N_{c}$ under the hidden gauge group $SU(N_{c})$. Briefly speaking, this scenario is just an easy extension of the SM by replacing the Higgs mass term with the hidden scalar QCD sector plus a Higgs portal coupling, which can overall be described by the following Lagrangian: 
\begin{eqnarray}
    {\cal L}  &=& - \frac{1}{2} {\rm tr} [G_{\mu \nu} G^{\mu \nu}] + (D_\mu S_i)^{\dag} (D^\mu S^i)\nonumber \\ 
    && + \lambda_{HS} (H^{\dag} H)(S^{\dag} _i S^i) - \lambda_H (H^{\dag} H)^2 \nonumber \\
    && -\lambda_{S1} (S^{\dag} _i S^i)(S^{\dag} _j S^j) -\lambda_{S2} (S^{\dag} _i S^j)(S^{\dag} _j S^i) \, ,
    \label{eq1}
\end{eqnarray}
where $D_\mu = \partial _\mu - i g_{\cal S} G_\mu$, with $g_{\cal S}$ being the coupling constant of the hidden scalar QCD, $H$ is a Higgs doublet having the 
same charge as the one in the SM, and the $G_\mu = \sum_{A=1}^{N_c^2-1} G_\mu ^A (t^A/2)$ correspond to the gauge fields coupled with the generators of $SU(N_c)$ group, $(t^A/2)$. 

At the classical level, the whole system possesses the hidden $SU(N_c)$ gauge symmetry and a $U(N_f)$ global symmetry, as well as the classical-scale invariance. Due to the asymptotic freedom of the strong dynamics~\footnote{
If we assume that these scalar fields interact in the fundamental representations, then the existence of the asymptotic freedom for the scalar QCD (at least one-loop level) implies that the numbers of colors and flavors are constrained to satisfy $11 N_c > N_f$.}, 
at a low-energy scale, the gauge coupling $g_{\cal S}$ grows up to be nonperturbatively large, driving the dynamical formation of an $U(N_f)$-invariant scalar-bilinear condensate~\footnote{Note that the Vafa-Witten theorem~\cite{Vafa:1983tf} protects the nonperturbative vacuum to be {\it parity} invariant, where the {\it parity} could definitely be assigned to the complex scalar $S_i$ when it is allowed to couple to the SM fermions via gauging the SM charges, for instance. Hence the condensation composed of odd number of scalars like $\langle S_i S_j S_k \rangle$ will not arise at the global minimum.}: 
\begin{equation}
    \langle (S_i ^\dag S_j) \rangle = \langle \sum_{a=1} ^{N_c} S_i ^{a \dag} S^{a} _j  \rangle \propto \delta _{ij} \, ,
    \label{SS-cond}
\end{equation}
which breaks the classical-scale invariance spontaneously. It also induces a negative Higgs mass term proportional to $\lambda_{HS}\langle S^\dag S \rangle$~\footnote{
As in the original scenario of the hidden scalar QCD~\cite{Kubo:2015cna} (and also in the original idea of EW symmetry breaking via the ordinary QCD with colored scalars~\cite{Kubo:2014ova}), the sign of the portal coupling $\lambda_{HS}$ (and the related $\lambda_{HM}$ in Eq.(\ref{Lag:LET}) and $\lambda_{H\chi}$ in Eq.(\ref{eq2})) has to be negative to trigger the Higgs Mechanism in the SM. However, unlike the SM, the classical scale invariance protects the hidden scalar QCD from a quadratic divergence even if we run the theory to the UV scale, and thus protects the model from the fine-tuning of the Higgs mass term.},  
to dynamically generate the EW scale ($v \simeq 246 \, {\rm GeV}$). In this sense, the origin of the EW-symmetry breaking and the classical-scale symmetry-breaking can be explained simultaneously by the hidden scalar QCD~\cite{Kubo:2015cna}. 

As in the case of the ordinary QCD, the color confinement will also be triggered dynamically, so that the low-energy dynamics will be governed by composite states like {\it scalar mesons and scalar baryons}. In the literature~\cite{Kubo:2015cna}, a low-energy effective model was proposed in a way analogously to the quark-meson model for the ordinary QCD, and the scalar QCD meson physics was discussed. There it was assumed that the generation of the scalar meson mass is dominated by the scalar condensate contribution as in Eq.(\ref{SS-cond}), so that the scale anomaly effect can be treated as a sub-leading order perturbation of the scale symmetry. 
However, as will be clearly demonstrated later, it turns out that the nonperturbative scale anomaly effect coming from the gluon condensation should also be incorporated as a crucial ingredient in employing a low-energy description for the hidden scalar QCD.

\subsection{Scale-invariant linear sigma model} 
As a low-energy description for the scale-invariant hidden scalar QCD with colored scalars in the fundamental representations, we can consider a linear sigma model constructed from the scalar mesons and baryons, denoted as $S_i ^\dag S_j \sim M_{ij}$ and $S_i S_j ^\dag S_k \sim B_{ijk}$, respectively, just like the fermionic QCD case. With the $U(N_f)$ global symmetry and the classical scale invariance, the most general Lagrangian can be written as follows:
\begin{eqnarray}
    {\cal L}_{\rm eff} &=& {\rm tr}[(\partial_\mu M)(\partial ^\mu M)] + (\partial_\mu B^\dag)(\partial ^\mu B) \nonumber \\
    &&- \lambda _{B1} B^\dag B ({\rm tr}[M^2]) - \lambda _{B2} B^\dag B ({\rm tr} [M])^2  \nonumber \\
    && -\lambda_{M1}({\rm tr}[M^4]) -\lambda_{M2}({\rm tr}[M^2])^2-\lambda_{M3}({\rm tr}[M])^4 \nonumber \\
    && +\lambda_{HM}(H^\dag H)({\rm tr}[M])^2  \, ,
\label{Lag:LET}
\end{eqnarray}
where it is assumed that all $\lambda$'s are positive, and the trace is taken over all flavor indices with the normalization for the $U(N_f)$ generators, ${\rm tr}(T_a T_b)=\frac{1}{2} \delta_{ab}$. 
After decomposing the field $M$ into the $U(N_f)$-flavor singlet and adjoint parts  
(denoted as $\chi$ and $A_0^a$, respectively) as $M_{ij} = \frac{1}{\sqrt{2N_f}} \chi \delta _{ij} +   A_0 ^a T_{ij} ^a $, we can redefine the coupling constants to write down the effective Lagrangian: 
\begin{eqnarray}
   {\cal L}_{\rm eff}  &=& - \lambda_H (H^\dag H)^2  
   - \lambda_\chi \chi ^4 -\lambda_{A_0} ^{abcd} A_0 ^a A_0 ^b  A_0 ^c A_0 ^d 
   \notag \\ 
   && -\lambda_{A_0 \chi} A_0 ^a A_{0} ^a \chi ^2  
     -\lambda_{B \chi} B^\dag B \chi ^2  -\lambda_{\chi A_0} ^{abc} \chi A_0 ^a A_0 ^b  A_0 ^c \nonumber \\
    && -\lambda_{B A_0} B^\dag B A_0 ^a A_{0} ^a  + \lambda_{H \chi}(H^\dag H)\chi ^2 
    + \cdots \, .
    \label{eq2}
\end{eqnarray}
As to the phenomenological consequence for the scalar baryon $B$ and $U(N_f)$-adjoint mesons $A_0$, we will make give some comments in the later section, and hereafter will focus on the singlet scalar $\chi$ coupled to the Higgs field $H$.

The target potential terms are then extracted as
\begin{eqnarray} 
    V(H, \chi) = -\lambda_{H\chi} (H^\dag H) \chi ^2 + \lambda_H (H^\dag H)^2 + \lambda_\chi \chi ^4 \, .
    \label{V-chi-H}
\end{eqnarray}
Defining the vacuum expectation values for $\chi$ and $H$  
as $\langle \chi \rangle = \eta$ and $\langle H \rangle = (0, v/\sqrt{2})^T$,  
 we assume this potential to reach it's minimum at the stationary condition:  
\begin{eqnarray}
  \frac{\partial V(v, \eta)}{\partial \eta} &=& 0 \,, \nonumber \\
  \frac{\partial V(v, \eta)}{\partial v} &=& 0 \, ,
    \label{eq4}
\end{eqnarray}
which has the solutions:
\begin{eqnarray}
    v ^2 &=& \frac{\lambda_{H\chi }}{\lambda_H} \eta ^2 \,, \nonumber \\
    \lambda_\chi &=& \frac{\lambda_{H\chi } ^{\quad 2}}{4 \lambda_H } \,.  
    \label{eq3}
\end{eqnarray}
Not surprisingly, the potential has a flat direction along both the $\eta$ and $v$ axes, where the vacuum expectation value of $\chi$ ($\eta$)  
generates the nonzero vacuum expectation value of the Higgs field $H$. In other words, we spontaneously break the EW symmetry as well as the scale symmetry simultaneously, just like what we have analyzed about the breaking structure in the underlying hidden scalar QCD in the previous arguments. 

Around the flat direction, we may find the physical spectra for scalars. To examine that, let us take the unitary gauge for the EW gauges, so that we can easily expand the potential by shifting fields like $\chi \to \eta + \chi $ and $H \to \frac{1}{\sqrt{2}}(0 , v+h)^T$,  
to get the mass terms: 
\begin{eqnarray}
     \frac{1}{2} m_h ^2 h^2 + \frac{1}{2} m_{\chi} ^2 \chi ^2 - 2 \lambda_{H \chi} 
     v \eta h\chi \, ,
     \label{eqm}
\end{eqnarray}
 with $m_h^2 =2\lambda_{H} v^2$ and $m_{\chi} ^2 =2\lambda_{H \chi} v^2$. 
 The last term reflects the mixing between $H$ and $\chi$. Actually, the determinant of this mixing matrix turns out to be zero, implying that a physical massless scalar boson. 
Thus a massless dilaton appears as the Nambu-Goldstone boson for the spontaneous-breaking {of} scale-invariance, known as {\it scalon}~\cite{Gildener:1976ih}. 
But we must note that actually the scale symmetry is merely an approximation and is assumed to be explicitly broken by the quantum corrections, which eventually produce a massive pseudo Nambu-Goldstone boson, just like pions in the ordinary QCD. 
As will be clarified later, a possible inclusion of the nonperturbative-explicit scale-symmetry breaking effect into this {\it toy model} will allow us to have a heavy enough massive composite dilaton. 

Actually, the nonperturbative-explicit breaking effect, i.e., the nonperturbative scale anomaly signaled by the gluon condensate, can be closely tied also with the scalar QCD hadron spectra: in~\cite{Iida:2006fe,Iida:2007qp,Iida:2008cq} the first lattice simulation was performed for $SU(3)_c$ scalar QCD, showing that all of the scalar-quark hadrons obtain large quantum corrections dominantly from gluons, i.e., the gluon condensate without the chiral symmetry breaking. 
The estimated mass was shown to be as large as the scale of (inverse of) the lattice spacing, i.e., the cutoff scale (say, the intrinsic scale of the hidden scalar QCD, $\Lambda_{\rm hQCD}$), which can be understood as the usual quadratically divergent quantum correction to the scalar mass, just like the Higgs 
in the SM. 
Besides, a non-perturbative analysis with the Schwinger-Dyson formalism was also carried out~\cite{Imai:2014tea,Imai:2014yxa}, where it was shown that the mass of diquark bound state (which acts as a colored scalar in scalar QCD) is dynamically generated from the nonperturbative-gluonic dressing-effect. 
These results support the nonperturbative gluonic effect to be one of the important missing keys for predicting the realistic scalar QCD hadron spectrum.

To properly incorporate such an important gluonic effect into effective models for scalar QCD as the main mass-scale generator, in the present study we shall adopt some recent idea, dubbed a chiral-scale effective model, which has been developed as a low-energy description of the fermionic QCD~\cite{Li:2016uzn}. 
As will be clarified below, this idea can properly realize the important gluonic contribution playing the role of the mass-scale generator through the nonperturbative-explicit breaking of the scale invariance~\footnote{
In addition to the chiral-scale symmetry structure, the color confinement effect of these strongly coupled scalar fields can also be important, especially for the vacuum structure and phase transition. However, to discuss color confinement, we need more assumptions as well as lattice results to show that the scalar QCD behaves exactly the same as the fermionic QCD, as was discussed in \cite{Kubo:2015cna}. 
One recent paper~\cite{Kubo:2018vdw} tried to include the Polyakov loop effect into the effective model for this hidden scalar QCD and then found a significant enhancement to the energy density of the gravitational waves background produced from a first-order scale phase transition. So just like the fermionic QCD, the effective model for a scalar QCD is also extremely complicated, where the inclusion of color confinement can also be important within some context. However, this interesting issue is beyond scope of our present study.}.


\subsection{Leading-order scale symmetry}
As we discussed above, it is essential to remove the strong constraint of the classical scale invariance to give a mass to the dilaton, and also to include a non-perturbative gluonic effect from the underlying dynamics. One possible access to achieve them is to introduce a small explicit breaking term characterized by a breaking parameter $a$ arising from the scale anomaly, which is conventionally known as the leading order scale symmetry (LOSS) in the fermionic QCD~\cite{Li:2016uzn}: 
\begin{eqnarray}
     {\cal L}_{\rm LOSS} = {\cal L}_{\rm eff} - \lambda_a \eta^4 \left( \frac{{\chi}}{\eta} \right) ^{4+a}  \,, \label{LOSS-LAG}
\end{eqnarray}
where ${\cal L}_{\rm eff}$ is given in Eq.(\ref{eq4}), and  
 $\lambda_a$ is a dimensionless coupling constant and has been normalized by the vacuum expectation value of dilaton ($\eta$). 
The explicit-breaking parameter $a (\neq 0)$ acts actually as an anomalous dimension for the gluon condensate, as will be seen later.  
This Lagrangian then no longer possesses the classical-scale invariance, but can still have an approximate scale symmetry if the explicit-breaking parameter $a$ is small enough. In that case, the scale symmetry is explicitly broken {\it minimally and only} at the dilaton potential, which is the reason why it is called the leading order scale symmetry {(LOSS)}.

To see how this explicit breaking term originates from the quantum gluonic corrections, we may only concentrate on the self interaction terms for the $\chi$ in Eq.(\ref{LOSS-LAG}), and derive the partially-conserved dilaton-current (PCDC) relation associated with the dilatation current $D_\mu = \theta _{\mu \nu} x^\nu$. 
So, now suppose we have a potential:
\begin{eqnarray}
    V_{\rm LOSS}(\chi) =   \lambda_\chi \chi^4 + \lambda_a \eta ^4 (\frac{\chi}{\eta}) ^{4+a} \, ,
\label{LOSS:potential}
\end{eqnarray}
then we can easily write down (the symmetric part of) the trace of the energy-momentum tensor that equals to the derivative of dilatation current:
\begin{eqnarray}
T_\mu ^\mu &=& \partial_\mu D^\mu = \delta_D {\cal L}= - (a \cdot \lambda_a) \cdot\eta ^4 \left( \frac{\chi}{\eta} \right) ^{4+a} \, ,
\label{Tmumu}
\end{eqnarray}
where $\delta_D$ is an infinitesimal variation under the scale transformation. 
Using the $\chi$-mass formula derived from the potential in Eq.(\ref{LOSS:potential}), $m_\chi ^2 = (a \cdot \lambda_a) (4+a) \eta^2$, we evaluate Eq.(\ref{Tmumu}) at the vacuum with $\chi = \eta$ as 
\begin{eqnarray}
\langle T_\mu ^\mu \rangle = - \frac{m_{\chi} ^2 \eta ^2}{4+a} \, .
\label{PCDC:LOSS}
\end{eqnarray}
This is the PCDC relation corresponding to the scale anomaly effect from the last term in Eq.(\ref{LOSS:potential}). 
Note from Eq.(\ref{PCDC:LOSS}) that the scale dimension of $T^\mu_\mu$ has been modified from the canonical dimension 4, by an anomalous dimension $a$ of the gluon condensate evaluated at some infrared scale, say, the hidden QCD scale 
$\Lambda_{\rm hQCD} = {\cal O}({\rm TeV})$: $d[T_\mu ^\mu]=d[(G_{\mu \nu})^2]=4+a$. Thus the parameter $a$ surely plays the role of the anomalous dimension of gluon condensate. 

\subsection{The LOSS model}

Now we are ready to incorporate the nonperturbative-scale anomaly effect 
into the linear sigma model by taking the LOSS limit. 
The $a$-anomalous dimension term in Eq.(\ref{LOSS:potential}) together with the Higgs potential and portal coupling parts  
in Eq.(\ref{V-chi-H}) then leads to a deformed flat direction:  
\begin{eqnarray}
    v ^2 &=& \frac{\lambda_{H\chi }}{\lambda_H} \eta ^2 \, , \nonumber \\
    \lambda_a + \lambda_\chi &=& \frac{\lambda_{H\chi } ^{\quad 2}}{4 \lambda_H }\, . 
\label{stationary}
\end{eqnarray}
Around this new vacuum, the physical Higgs and dilaton fields ($h', \chi'$) arise 
through the mixing structure similar to Eq.(\ref{eqm}) as  
\begin{eqnarray}
     \chi ' &=& \chi \cos{\theta} - h \sin{\theta} \nonumber \\
      h' &=& \chi \sin \theta + h \cos \theta  
      \,. \label{diag} 
\end{eqnarray} 
Then one finds the masses of the $h'$ and the $\chi '$ expressed like   
\begin{eqnarray}
m_{h'}^2  &=& 2 
\left( \frac{\cos ^2 \theta}{\cos{2\theta}} \lambda_{H} - \frac{\sin^2 \theta}{\cos{2\theta}} \lambda_{H \chi} \right) \cdot v ^2 
\,, \notag \\ 
   m_{\chi '} ^2 &=&2\left( \frac{\cos ^2 \theta}{\cos{2\theta}} \lambda_{H\chi} - \frac{\sin ^2 \theta}{\cos{2\theta}} \lambda_{H} \right) \cdot v ^2 \nonumber \\
    &&+ 7 \left( \frac{\cos ^2 \theta}{\cos{2\theta}} 
    \left( a\cdot \lambda_a \right) \frac{\lambda_H }{\lambda_{H\chi}} \right) \cdot v^2 \nonumber \\
    &\equiv & 2\left( \frac{\cos ^2 \theta}{\cos{2\theta}} \lambda_{H\chi} - \frac{\sin ^2 \theta}{\cos{2\theta}} \lambda_{H} \right) \cdot v ^2 + \Delta m_{\chi '} ^2  
    \label{Delta-mchi-prime}
\end{eqnarray}
where we have assumed that the explicit-breaking effect should be small enough 
to preserve an approximate scale invariance, so that terms coming with the anomalous dimension $a$ can be expanded as $\chi ^a =1+a \ln \chi + {\cal O}(a^2)$.  
Note that the $h'$ mass has no $a$-dependence, while the $\chi'$-dilaton gets 
corrections of ${\cal O}(a)$. The latter term reflects the effect of LOSS as discussed in the previous section, and for later convenience will be specifically defined as $\Delta m_{\chi '} ^2$ in the last line of Eq.(\ref{Delta-mchi-prime}). 
 
To demonstrate the crucial gluonic effect at the LOSS limit, 
here we shall take a simple-minded ansatz: assume the dilaton mass is {around} the vacuum expectation value $\eta$. A similar situation can be seen in the current algebra argument applied to the ordinary QCD pion, arising as the pseudo Nambu-Goldstone boson for the spontaneous-chiral symmetry breaking, where the pion mass scale is nearly identical to the decay constant scale $f_{\pi}$. {Thus} in the ideal limit we may set $\eta = m_{\chi '} $. 
The physical Higgs mass $(m_{h'})$ and the EW scale $(v)$ are fixed to be $125$ GeV and $246$ GeV, respectively. As to the mixing angle $\theta$, since the couplings of the 125 GeV Higgs $h'$ to SM particles scale overall by $\cos\theta$ compared to the SM prediction and also all new particles predicted from the present hidden QCD do not carry any SM charges, it can be set as $\cos ^2 \theta=0.9$ by referring to the current upper bound on the Higgs coupling measurement~\cite{Aad:2015pla}. 
Then, taking $a=0.1$ as a reference point in order to realize the approximate scale invariance, we can numerically estimate the dilaton mass and values of other couplings to find:
\begin{eqnarray}
    &&\eta \equiv m_{\chi '}   \simeq  282 \, {\rm GeV} \,,\notag \\ 
    && \lambda_{H\chi}  \simeq  0.138 \,, \qquad 
    \lambda_H  \simeq  0.182  \,, \notag \\ 
    && \lambda_a  \simeq  1.014 \,, \qquad 
    \lambda_\chi  \simeq  -0.987 \, .  
    \label{values}
\end{eqnarray} 
From this parameter set we have $\Delta m_{\chi '} ^2  \simeq  1.051 v^2$, hence  
\begin{eqnarray} 
\frac{\Delta m_{\chi '} ^2}{ m_{\chi '} ^ 2}  \simeq  0.798 \,. 
\end{eqnarray} 
This clearly shows that the mass of the $\chi'$-dilaton (mostly composed of scalar bilinear) is dominantly supplied by the gluon condensate. This is 
in agreement with the results from the lattice simulation for the scalar QCD 
and some other related nonperturbative computations~\cite{Iida:2006fe,Iida:2007qp,Iida:2008cq,Imai:2014tea,Imai:2014yxa}, even though the composite scalar mass ($\sim 280$ GeV) is smaller than the order of the inverse of lattice spacing (i.e. the cutoff scale $\sim \Lambda_{\rm hQCD} = {\cal O} ({\rm TeV})$) due to the assumed small-anomalous dimension $a$ at the LOSS limit.

Before closing this section, we emphasize that 
as long as the value of $a$ is small enough ($a \ll 1$)  
the scale-anomaly effect arising from the perturbative expansion as 
$\chi ^a =1+a \ln \chi + {\cal O}(a^2)$ 
always takes a combined form as $(a \cdot \lambda_a)$ (up to ${\cal O}(a^2)$) when entering into all the effective couplings (with making use of the stationary condition in Eq.(\ref{stationary})). Therefore, the mass of $\chi'$ dilaton is fixed to the number in Eq.(\ref{values}) whatever the anomalous dimension $a$ has been taken. It also implies that even in evaluating cross sections involving the dilaton $\chi'$, only the combination of $(a \cdot \lambda_a)$ can be observed and determined from the experiment. Once we fix the mixing angle $\theta$, the factor $(a \cdot \lambda_a)$ will also be fixed to a constant automatically. So in practice, the phenomenology of this model will not be affected by the value we choose for the small anomalous dimension $a$, which makes this model more predictive.


\section{Phenomenological predictions from the LOSS} 
In this section, we discuss the specific phenomenological predictions on the 
light $\chi'$-dilaton at the LHC. 

\subsection{Decay property of the $\chi'$-dilaton}
We first note that
the original composite dilaton $\chi$ is totally SM singlet, hence 
initially has no coupling to fermions and gauge bosons in the SM. 
The $\chi'$ couplings, which can be sensitive to experiments, arise only  
due to the mixing with the original Higgs $H$, which should be small enough 
to be consistent with the current Higgs coupling measurement   
(i.e. $\sin^2 \theta \le 0.10$~\cite{Aad:2015pla}).  
Nevertheless, the $\chi'$-dilaton with the mass around 
280 GeV (as in Eq.(\ref{values})) can have a chance to 
be generated at the LHC through the same process as we produce the SM-like Higgs, 
which would be dominated by the gluon-gluon fusion production. 
(The scale anomaly arising from the EW interactions actually provides the leading-order $\chi'$-dilaton coupling to diphoton and digluon 
(just like the SM Higgs case), which comes 
from the mixing with the SM-like Higgs via the Higgs portal coupling.) 
Similarly, the $\chi'$-dilaton decays into the SM particles in the same way as the SM-like 
Higgs with the mass around 280 GeV, 
where the dominant decay channels are expected to be  
signaled by $WW$, $ZZ$, and the diHiggs modes (if $m_{\chi'} > 2 m_{h'}$). 

\begin{figure}[htbp]
\begin{center}
    \vspace{0.5cm}
    \includegraphics[width=0.40\textwidth]{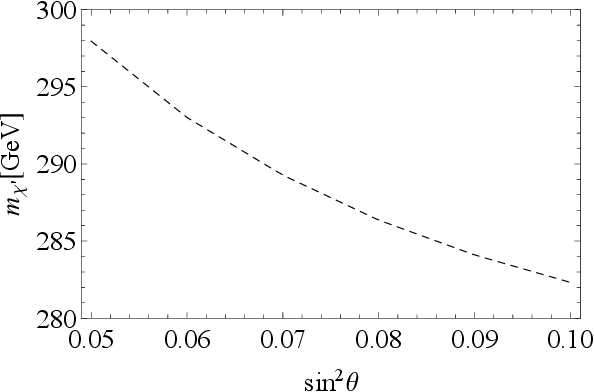}
    \caption{The physical dilaton ($\chi'$) mass as a function of $\sin^2 \theta$}
    \label{fig1}
\end{center}
\end{figure}

Now, let us allow the mixing strength, $\sin ^2 \theta$, to vary from the current bound (0.1) down to 0.05, in light of the prospected reach of the high-luminosity LHC (HL-LHC) for the Higgs coupling measurement~\cite{Dawson:2013bba}. 
In the ideal limit following the simple-minded ansatz we have made 
($m_{\chi'} \equiv \eta$),  the $\chi'$ mass $m_{\chi'}$ is estimated to be given as a function of $\sin^2\theta$, as shown in Figure \ref{fig1}. 
There we see that the $\chi'$ mass is necessarily greater than twice of the Higgs mass, which kinematically allows the decay into diHiggs. 

To discuss the $\chi'$ decay property, we should also study the trilinear couplings among the $\chi '$ and the $h'$. The definitions for the couplings we use are: 
\begin{eqnarray}
{\cal L}_{\rm trilinear} &=& - \frac{1}{3!} 
 \lambda_{h'h'h'} {h' }^3 
- \frac{1}{3!} 
\lambda_{\chi' \chi' \chi'} {\chi' }^3  \nonumber \\
&&-  \frac{1}{2!} 
\lambda_{h'h'\chi'} {h'}^2 \chi ' 
- \frac{1}{2!} 
\lambda_{h' \chi' \chi'} h' {\chi ' }^2 \, .
\label{hhhc}
\end{eqnarray} 
Note that for these trilinear couplings, the sign of trigonometric function becomes important. In particular, one can easily check that the $\lambda_{h'h' \chi}$ coupling relevant to the $\chi' \to h'h'$ decay is expressed by the expansion in powers of the anomalous-dimension parameter $a$: 
\begin{eqnarray}
\lambda_{h'h'\chi'}
&=& 2 \cdot \Bigg\{ 12 \cdot \left( \lambda_\chi +\lambda_a \left( 1+\frac{13}{12} \right) a \right) \cdot \eta \sin ^2 \theta \cos\theta \nonumber \\
&&+ 3 \lambda_{H} v \sin \theta \cos ^2 \theta \nonumber \\
&&+ 2 \lambda_{H\chi} \sin \theta \cos \theta ( v\cos \theta +\eta \sin \theta ) \nonumber \\
&&- \lambda_{H\chi} \left( v\sin^3 \theta +\eta \cos^3 \theta \right) \Bigg\} \,, 
\end{eqnarray}  
with terms of ${\cal O}(a^2)$ omitted. 
Therefore, by varying the input mixing angle, 
the partial width for $\chi' \to h'h'$ as well as 
the total decay width ($\Gamma _\chi$) will have an explicit dependence not only on the $\sin^2 \theta$, but also on the sign of $\cos \theta$. 

\begin{figure}[htbp]
\begin{center}
    \vspace{0.5cm}
    \includegraphics[width=0.40\textwidth]{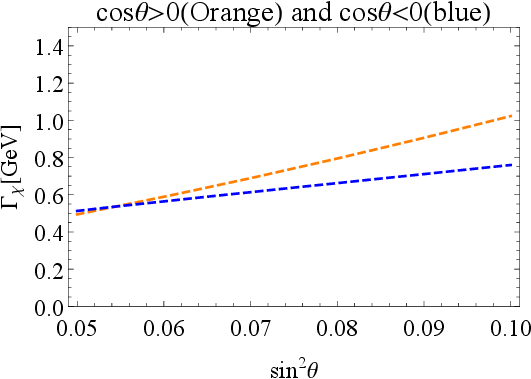}
    \caption{The $\chi'$-total decay width $\Gamma _\chi$ 
    as a function of $\sin^2\theta$, in unit of GeV.}
    \label{fig2}
\end{center}
\end{figure}

\begin{figure}[htbp]
\begin{center}
    \includegraphics[width=0.40\textwidth]{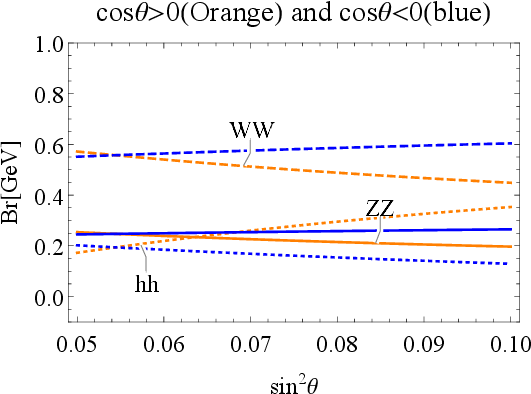}
    \caption{The $\chi'$-branching ratios for 
    the dominant $WW$ (dashed curves), $h'h'$ (dotted curves), and $ZZ$ (solid curves) decay modes, as a function of the mixing angle with the sign ambiguity by $\cos\theta$.}
    \label{fig3}
\end{center} 
\end{figure}

Figure \ref{fig2} shows the total width $\Gamma_\chi$ plotted as a function of $\sin^2 \theta$ for $\cos\theta >0$ and $\cos\theta<0$. 
Figure~\ref{fig2} in combination with Fig.~\ref{fig1}
implies that the $\chi'$-dilaton is a sufficiently narrow resonance 
having the width-to-mass ratio around $10^{-3} - 10^{-2}$.   
In Table~\ref{table1} we list the estimated 
numbers for the $\chi'$-branching ratios together with the total width for some particular choices for the angle $\theta$ including the sign ambiguity. 
In Figure \ref{fig3}, we plot the branching ratios for the dominant $WW$, $hh$, and $ZZ$ decay modes as a function of the mixing angle.
\begin{table*}[!htbp] 
\centering
\begin{tabular}{|c|c|c|c|c|c|c|c|c|}
\hline
$\sin^2 \theta$ & $\cos \theta$ & $m_{\chi '} [\rm{GeV}]$ & $\Gamma_\chi [\rm{GeV}]$ & Br($WW$) [\%] & Br($hh$) [\%] & Br($ZZ$) [\%]& Br($b \bar{b}$) [\%] & Br($\gamma \gamma$) [\%] 
\\ \hline
0.1 & + & 282 & 1.023 & 44.9  & 35.3 & 19.7  & 0.095  & 0.00071  \\ \hline
0.1 & - & 282 & 0.760 & 60.4  & 12.9  & 26.5  & 0.128  & 0.00095  \\ \hline
0.05 & + & 298 & 0.494 & 57.2  & 17.3  & 25.4  & 0.103  & 0.00083  \\ \hline
0.05 & - & 298 & 0.513 & 55.1  & 20.2  & 24.5  & 0.100  & 0.00080  \\ 
\hline
\end{tabular}
\caption{Numerical results for total decay width and branching ratios of the $\chi'$-dilaton, with the mixing angle varied}.
\label{table1}
\end{table*}

\subsection{Effect on the Higgs trilinear coupling}
Similarly to the $\lambda_{h'h'\chi'}$ coupling, 
the Higgs trilinear coupling $\lambda_{h'h'h'}$ 
in Eq.(\ref{hhhc}) explicitly depends on the sign of mixing angle as well as 
the anomalous-dimension parameter $a$: 
\begin{align} 
\lambda_{h' h'h'} 
= &
- 4 \sin^3 \theta \left[\lambda_\chi + \left(1 + \frac{13}{12} a \right) 
\lambda_a  \right] \eta  
\notag \\  
& +
\cos^3\theta \lambda_H v  
\notag \\ 
& - \sin \theta \cos \theta  \lambda_{H \chi} 
  \left( v \sin \theta - 
  \eta \cos \theta \right)
\,, 
\end{align}
up to terms of ${\cal O}(a^2)$. It turns out that the present LOSS model (with nonzero $a$) can provide a significant difference from the SM's prediction, as well as the prediction from the classical scale-invariant model (with $a=0$).  
For simplicity, we may take $\sin ^2 \theta =0.1$ and compare the Higgs trilinear coupling for these three models in the unit of the EW scale $v$.  
In the SM, the straightforward calculation yields $\frac{\lambda_{hhh}}{v}|_{\rm SM} \simeq 0.75$. 
In the classical scale-invariant hidden scalar QCD (i.e. $a=0$), 
$\frac{\lambda_{h'h'h'}}{v}|_{a=0} \simeq 0.652$ 
for $\cos \theta > 0$ and 
$\frac{\lambda_{h'h'h'}}{v}|_{a=0}\simeq-0.522$
for $\cos \theta <0$. 
So, the Higgs trilinear coupling is totally 
decreased by about 13\% or 30\% with respect to the SM's value. 
Including the LOSS correction with $a=0.1$, we find 
$\frac{\lambda_{h'h'h'}}{v}|_{\rm LOSS}\simeq 1.01$
for $\cos \theta >0$ and 
$\frac{\lambda_{h'h'h'}}{v}|_{\rm LOSS}\simeq-0.702$
for $\cos \theta <0$. 
Therefore, we can either maximally have about 34\% enhancement to the SM's value, or have a nearly intact Higgs trilinear coupling in magnitude. 

Therefore, in addition to the predicted $\chi'$-dilaton spectrum and its coupling property, from the difference of the Higgs trilinear coupling we may note the significance of including the nonperturbative scale-anomaly effect from the hidden gluons condensate. This would allow us to emphasize that if any deviation regarding the Higgs trilinear coupling can be found in the future collider experiments, then it could be a significant signal for probing the hidden scalar QCD at the LOSS limit.

\subsection{Predicted diHiggs signals}
Now we discuss the predicted diHiggs production cross sections at the LHC including the $\chi'$-dilaton exchange contribution at the LOSS limit. The diHiggs signals are dominantly generated by the gluon-gluon fusion through the top-quark box diagram and the top-quark triangle diagram involving the Higgs or the dilaton exchange. 
To see how the $\chi'$ resonance contributes, we shall just take the heavy top-quark limit ($m_t \to \infty$) to show the analytic expression for the total diHiggs cross section at the leading order (LO)in the QCD perturbation\footnote{The finite top mass effects at LO will be taken into account later.}:   
\begin{eqnarray}
&& \sigma (gg \to h'h')\Bigg|^{\rm LO}_{m_t \to \infty} \notag \\ 
&&= 
\frac{\alpha_s ^2}{512(2 \pi) ^3 v^2}\cdot  b_t ^2 
\cdot \hat{s} \sqrt{\frac{\hat{s}-4(m_{h'})^2}{\hat{s}}} \nonumber \\
&&  \times \left| \frac{1}{v} \cos ^2 \theta 
-\frac{ \lambda_{h'h'h'} \cos \theta}{\hat{s}-(m_{h'})^2 + i m_{h'} \Gamma_{h'}} \right. \nonumber \\
&& \left. - \frac{ {\lambda_{h'h'\chi'}} \sin \theta}{\hat{s}-(m_{\chi '})^2 + i m_{\chi '} \Gamma_{\chi'}} \right| ^2 \, ,
\label{cross-diHiggs}
\end{eqnarray}
where $b_t = \frac{2}{3}$ given in the heavy top-quark limit, $\alpha_s \simeq 0.118$ which denotes the fine-structure constant for the QCD gauge 
coupling evaluated at the EW ($Z$ boson mass) scale~\cite{Tanabashi:2018oca}, 
$\hat{s}$ is the total center of mass energy at the parton level, and $\Gamma_h \simeq 0.0041$ GeV is the total decay width of the Higgs boson predicted 
in the SM~\cite{Denner:2011mq}. As it will turn out, the diHiggs cross section is almost (more than 80\% in magnitude) 
saturated by the $\chi'$-resonant term (the last term in Eq.(\ref{cross-diHiggs})), 
mainly because the predicted $\chi'$ mass (as in Fig.~\ref{fig1}) is quite close to the threshold for the on-shell diHiggs production along with a 
sizable enough effective-trilinear coupling $(\sin\theta \cdot \lambda_{h'h'\chi'})$.

The diHiggs cross section via the gluon-gluon fusion process 
at the LHC is evaluated by the convolution integral as  
\begin{align}
& 
\sigma (pp \to h'h')= \notag \\ 
& 
\int_{\tau_0}^1 dx_1 \int_{\tau_0/x_1}^1 dx_2 \, f_{g/p}(x_1,\mu_F) f_{g/p}(x_2,\mu_F) 
\sigma (gg \to h'h') \, ,
\label{convol}
\end{align}
where $f_{g/p}$ denotes the parton distribution function (PDF) for gluon parton in the proton and $x_{1,2}$ are the Bjorken variables relating the $\hat{s}$ {to} the center of mass energy for the proton-proton collision, $s$, like $\hat{s}= x_1 x_2 s$. The kinematical cutoff parameter $\tau_0$ (normalized by $s$) will be set to the threshold energy for the diHiggs production: $\tau_0 = (2 m_h)^2/s$, and the factorization scale $\mu_F$ will be taken as $\mu_F = \sqrt{\hat{s}}$. 

To get a rough estimation on the $\chi'$ contribution to the diHiggs cross section, we first attempt to work on the LO estimation at the heavy top-quark limit by simply setting the parton-level cross section $\sigma(gg \to h'h')$ in Eq.(\ref{convol}) to the one displayed in Eq.(\ref{cross-diHiggs}).
In that case, the convolution integration in Eq.(\ref{convol}) is easily 
computed by using the {\tt CUBA} package~\cite{Hahn:2004fe}  
implementing the {\tt CTEQ6L1} parton distribution function~\cite{Pumplin:2002vw} in {\tt Mathematica} with the help of a PDF parser package, {\tt ManeParse\_2.0}~\cite{Clark:2016jgm}. By varying the value of $\sin^2 \theta$ in the range from 0.05 to 0.1, we can thus numerically evaluate the diHiggs cross section at the LHC with the center of mass energy $\sqrt{s}$ set to 13 TeV, 
to find that the $\chi'$ contribution from the last term in Eq.(\ref{cross-diHiggs}) is highly dominant to saturate the total cross section by about $(78\%, 92\%)$ for $\sin^2 \theta =(0.10, 0.05)$ with $\cos\theta >0$, and $(99\%, 93\%)$ for $\sin^2 \theta =(0.10, 0.05)$ with $\cos\theta <0$. This remarkable dominance arises because the on-shell $\chi'$ production with the narrow width as in Fig.2 has been realized, and also due to the threshold enhancement for the on-shell diHiggs production by the $\chi'$ resonance with a large effective trilinear coupling $(\sin \theta \cdot \lambda_{h'h'\chi'})$ comparable to the effective Higgs trilinear coupling $(\cos \theta\cdot \lambda_{h'h'h})$, which overwhelms the destructive interference with the top-box graph, though the latter also gets somewhat more stringent than the SM case. Thus, just to demonstrate the essential feature of the present LOSS prediction for the diHiggs event at the LHC, it would be sufficient to approximate the diHiggs cross section just by the $\chi'$ resonant part. 

\begin{figure}[htbp]
\begin{center} 
    \vspace{0.5cm}
    \includegraphics[width=0.40\textwidth]{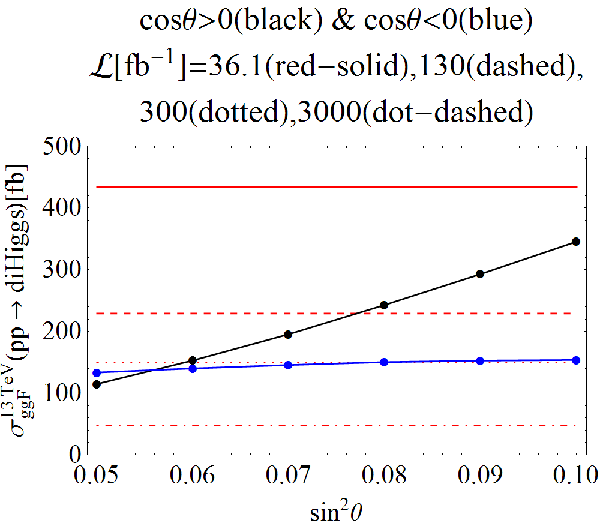}
    \caption{The predicted diHiggs production cross section $\sigma_{ggF} ^{13 \rm{TeV}} (pp \to h'h')$ as a function of $\sin ^2 \theta$. 
    The red horizontal lines are experimental upper bounds coming from 
    non-resonant production, which have been scaled from the currently most 
    stringent limit placed by the ATLAS group~\cite{Aaboud:2018knk} to 
    a various of luminosities to be accumulated in the near future,  
    as described in the plot label and in the text.}
    \label{fig4}
\end{center}
\end{figure}

To be more accurate, it is also necessary to take the finite top mass effects into account at LO. Since the diHiggs production is highly dominated by the on-shell $\chi'$ production decaying to diHiggs~\footnote{Even when the finite top mass effect is included, the on-shell narrow-$\chi'$ production term is roughly larger by factors of $(m_{\chi'}/\Gamma_{\chi})^2 \sim 10^4$ and $(m_{\chi'}/\Gamma_{\chi}) \sim 10^2$ than the off-shell Higgs resonant term as well as the non-resonant box terms, and the interference term, respectively, in the magnitude of the cross section.}, we may simply estimate the finite top mass correction to Eq.(21), by multiplying the scale factor like:
\begin{eqnarray} 
\frac{b_t^2(m_t)}{b_t^2} \simeq  1.5 (1.6) 
\,, \qquad {\rm for} \quad \sin^2 \theta =0.1(0.05)
\,,  \label{scale-fac}
\end{eqnarray} 
where $b_t(m_t) = (\sin^{-1} [\frac{m_{\chi'}}{2m_t}])^2$ arises from the top-quark triangle loop function. Including this scaling factor, we can roughly estimate the diHiggs cross sections at the LHC with $\sqrt{s}=13\, \rm{TeV}$ as a function of $\sin^2\theta$, as shown in Fig.~\ref{fig4}. The $\sin^2\theta$, ranged from 0.1 down to 0.05, would monitor the currently accumulated luminosity relevant for the diHiggs event with ${\cal L}=36.1\,{\rm fb}^{-1}$, up to the prospected one realized at the HL-LHC with ${\cal L}=3000\,{\rm fb}^{-1}$~\cite{Dawson:2013bba}. 
In the figure, we have simply multiplied by the K-factor for the $gg$-$hh$ type, $K\simeq 1.9$~\cite{Dawson:1998py}, which comes from the next-to leading order (NLO) QCD corrections to the diHiggs production in the heavy top-quark limit. The figure also shows that the predicted diHiggs cross section are significantly larger than the SM's prediction. For instance, by allowing the maximal mixing strength at present, i.e., taking $\sin ^2 \theta =0.1$ and $\cos\theta>0$, we have:
\begin{eqnarray} 
\sigma_{ggF} ^{13 \rm{TeV}} (pp \to h'h') \Bigg|_{\sin^2\theta =0.1,\, \cos \theta >0} \simeq 346 \, {\rm fb}\, , 
\label{benchmark}
\end{eqnarray} 
which is approximately 10 times larger than the SM prediction. Again note that this enhancement happens because of the threshold effect triggered by the $\chi'$ resonant contribution with the mass around 300 GeV.

The latest and most stringent upper bound (at 95\% C.L.) for the non-resonant diHiggs production cross section comes from the diHiggs to $b\bar{b} b\bar{b}$ channel in the ATLAS experiment with 36.1 $\rm{fb}^{-1}$ \cite{Aaboud:2018knk}, which is about 434 fb (13 times the SM expectation). This bound has been incorporated as the red-solid horizontal line in Fig.~\ref{fig4}. Just by naively scaling the current upper bound with respect to the luminosity, we can see from the figure that all the LOSS predictions can be probed at the luminosity $\sim$ 300 $\rm{fb}^{-1}$. 
In short, if any of the enhancement for the non-resonant diHiggs cross section is to be found in the future, then it would be a strong supporting signal for the LOSS, which can be explored at the HL-LHC with a high sensitivity. 
\begin{table*}[htb] 
\centering
\begin{tabular}{|c|c|c|c|c|c|c|c|}
\hline
$\sin^2 \theta$ & $\cos \theta$ & $m_{\chi '}$ [GeV] & $\sigma$($\chi '\to \gamma \gamma $) [fb] & $\sigma$($\chi '\to ZZ$) [fb] &$\sigma$($\chi '\to WW$) [fb] & $\sigma$($\chi ' \to h'h' \to b\bar{b}b\bar{b}$) [fb] \\ \hline
0.1 & + & \, 282  \, & 
0.0066  & 158  & 361  & 117   \\ \hline
0.1 & - & \, 282  \, & 
0.0088  & 213  & 486  & 52.1  \\ \hline
0.05 & + & \, 298  \, & 
0.0037  & 97.9  & 220  & 39.0  \\ \hline
0.05 & - & \, 298 \, & 
0.0035  & 94.4  & 213  & 45.2  \\ 
\hline
\end{tabular}
\caption{
Numerical results for the $\chi'$-dilaton resonant cross sections for different final states. 
The numbers for the 4b signals through the diHiggs production process 
have also been listed by referring to the diHiggs-cross section values 
in Fig.~\ref{fig4} with the branching fraction to $b\bar{b}$ for the SM Higgs multiplied 
twice.}   
\label{table2}
\end{table*}

At last but not least, as has been clarified recently in the literature~\cite{Borowka:2016ehy,Borowka:2016ypz,Baglio:2018lrj}, 
the full top-mass effect turns out to be significant for the diHiggs cross section including the NLO corrections. It would make the $K$ factor nonuniform with respect to  
the mass distribution function in the invariant mass range (from 300 GeV to 1000 GeV), in contrast to a rough overall multiplication as in the above analysis, and would eventually make a somewhat large deviation from the previous analysis, which has been reported to reach a 
maximally about 50\% level~\cite{Borowka:2016ehy,Borowka:2016ypz,Baglio:2018lrj} in the relevant differential distribution. Simply applying those 
full-top mass dependence in~\cite{Borowka:2016ehy,Borowka:2016ypz,Baglio:2018lrj} to our prediction, and taking into account the high-$\chi'$ resonant dominance around $280 - 300$ GeV in the mass distribution function, we may suspect that as far as the order of magnitude on the cross section is concerned, such a large top-mass dependence would be negligible: since our predicted cross section is sharply peaked at around 280 or 300 GeV with the small enough width $\lesssim 1$ GeV (as seen from Fig.~\ref{fig2}) where the total amount of the diHiggs rate has been saturated by $\gtrsim 80$\%, the $K$-factor will keep staying as a constant to be $K\simeq 1.9$~\cite{Borowka:2016ehy,Borowka:2016ypz,Baglio:2018lrj}.
Thus, the LOSS predictions shown in Fig.~\ref{fig4}, including the benchmark number in Eq.(\ref{benchmark}), would be intact even when the full-top quark mass dependence is incorporated, as long as we focus on the order of magnitude which suffices to our central claim on the importance of the anomalous gluonic effect on the hidden QCD-like scenario at the LOSS limit.

\subsection{Other $\chi'$-resonant signatures} 
Finally, we discuss the resonant dilaton productions in several channels 
at the LHC. 
Because the $\chi'$-total width is small enough as seen from Table~\ref{table1} 
or Fig.~\ref{fig2}, 
the $\chi'$-resonant cross section for the production of AB particles in the final state can be evaluated safely by using the narrow width approximation (NWA):
\begin{eqnarray}
\sigma (pp \to \chi ' \to AB) |_{\rm NWA}
 = 
\frac{\pi ^2}{8 \hat{s}} \cdot   \frac{ \Gamma (\chi ' \to gg) }{m_{\chi '} \rm{Br}(AB)} &&
\notag\\ 
\times \int_{-Y_B} ^{Y_B} d Y \, f_{g/p}(\frac{m_{\chi'}}{\sqrt{s}} e^{Y} , m_{\chi '}^2) f_{g/p}(\frac{m_{\chi'}}{\sqrt{s}} e^{-Y} , m_{\chi '}^2) \, , &&
\end{eqnarray} 
where the $K$ factor is read as 
$K\simeq 1.76$ (corresponding to the $gg$-$h$ production type 
at the resonant mass around 300 GeV)~\cite{Djouadi:2005gi,19,Dawson:1998py}, 
and $Y$ is the rapidity of the A-B final state system, cutoff by the pseudo rapidity $Y_B = -\frac{1}{2}\ln ({m_{\chi '} ^2}/{s})$. 
Table~\ref{table2} shows the values of several resonant cross sections, 
including 
$\chi '\to \gamma \gamma$, $\chi '\to ZZ$, $\chi '\to WW$ signals, 
and also $\chi '\to b\bar{b}b\bar{b}$ via the diHiggs production. 
The table tells us that all the predicted cross section signals in the present LOSS model are still somewhat smaller than the current upper bounds from the LHC experiments~\cite{Aaboud:2018knk,Aaboud:2017yyg,Aaboud:2018bun}, except the $WW$ plus $ZZ$ channel for $\sin^2 \theta =0.05$, where the upper bound on the cross section has currently been placed to be about 300 fb when the resonant mass is 300 GeV~\cite{Aaboud:2018bun}, close to or below the predicted numbers listed in Table~\ref{table2}. 

In addition to the sensitive $WW$ and $ZZ$ channels, 
more interestingly, recently the ATLAS group searching for diHiggs to 4b signals reported an excess for the resonant diHiggs production at around 280 GeV with a global significance of 2.3$\sigma$~\cite{Aaboud:2018knk}. 
In any case, it is expected that with increase of the luminosity and improvement more on the detectability for the diHiggs signal as well as other channels as listed in Table~\ref{table2}, the present LOSS model would be probed through some nontrivial resonant signals at the invariant mass around 280 GeV.

\section{Summary and Future Prospect}
In conclusion, we constructed an effective model for a hidden scalar QCD at the leading-order scale-symmetry (LOSS) limit, which properly includes a crucial gluon-condensate effect as the nonperturbative-scale anomaly term, and is in accordance with the indications from the lattice simulation and other nonperturbative analysis of the scalar QCD. Remarkably, it turned out that the LOSS model predicts a massive enough dilaton (with the mass around 280 GeV) as well as a number of detectable cross sections. This effective model is significantly different than the conventional realization of scale-invariant hidden-scalar QCD without the scale anomaly effect. The difference is manifested especially in the Higgs-trilinear coupling-term, the enhancement of LHC diHiggs production cross section (at most about 10 time larger than the SM prediction), as well as the other dilaton-resonant cross sections. In this sense, our model can potentially provide a competitive explanation for many exotic signals beyond the SM, which can be directly tested, or presumably be excluded by the future collider experiments, such as the high luminosity-LHC.

Finally, we want to emphasize that the existence of the rest of hidden QCD-composite scalars, $B$ and $A_0$ {as seen in Eq.(\ref{eq2})} might potentially provide us with some new multi-component DM scenario. Moreover, the large enough portal coupling $\lambda_{H\chi} \simeq 0.138$ (see Eq.(\ref{values})) also provides a possibility for a strongly first-order {EW} phase transition in the early universe. We leave a thorough discussion of such multicomponent DM scenario as well as the the consequences for the EW phase transition for future work.
\acknowledgments 
We would like to thank Kimmo Tuominen and Andi Hektor for careful reading of the manuscript and insightful comments, and Kenji Nishiwaki for several useful comments on the diHiggs phenomenology. We are also grateful to Matti Heikinheimo and Martti Raidal for useful comments in private discussion. 
S.M. is supported in part by the JSPS Grant-in-Aid for Young Scientists (B) No. 15K17645, National Science Foundation of China (NSFC) under Grant No.  11747308, and the Seeds Funding of Jilin University. R.O. was partially supported by the TAQ Honor Programs in physics from the Office of Undergraduate Education and College of Physics in Jilin University.


\end{document}